\documentstyle{elsart}

\begin{document}
\begin{frontmatter}

\rightline{DPNU-99-08}
\rightline{Mar. 9, 1999}
\bigskip

\title{Time-of-Propagation Cherenkov counter for particle identification} 

\author{M. Akatsu, M. Aoki, K. Fujimoto, Y. Higashino, M. Hirose,}\\
	{K. Inami, A. Ishikawa, T. Matsumoto, K. Misono, I. Nagai,}\\
	{T. Ohshima, A. Sugi, A. Sugiyama, S. Suzuki, M. Tomoto}
\author[INS]{and H. Okuno}
\address{Physics Department, Nagoya University \\
	Chikusa, Furo, Nagoya 464-8602, Japan}
\address[INS]{IPNS, High Energy Accelerator Research
	Organization (KEK), \\
	Ibaraki 305-0801, Japan}

\begin{abstract}
We describe here a new concept of a Cherenkov detector for 
particle identification by means of measuring the
Time-of-Propagation (TOP) of Cherenkov photons. 
\end{abstract}
\end{frontmatter}


\newpage

\section{Introduction}
\par
Particle-identification (PID) capability plays an essential role 
in experiments at B-factories. 
Especially, $\pi/$K identification in the momentum range up to 4 GeV/c 
over a wide angular region is crucially important for the primary
physics goals to measure CP violation. 
In the Belle detector at KEK-B, a combination of Aerogel Cherenkov 
counters, TOF counters and dE/dX measurement in a central drift 
chamber (CDC) provides PID information for charged particles [1]. 
Although the present detector system covers most of the momentum
region of the charged particles, the PID power for a 
particle with momentum above 3 GeV/c is not sufficiently satisfactory. 
Concerning the future upgrade of PID devices, we discuss here the
effectiveness of measuring both the Time-of-Propagation (TOP) and
the emission-angle of the Cherenkov photons.

The time resolution of the Time-of-Flight (TOF) counter using 
a plastic scintillator is inherently limited by the following effects 
besides the transit-time spread (TTS) of a phototube:
(1) a finite decay-time of photon emission, and (2) a 
different photon propagation-length or propagation-time in the 
scintillator to a phototube, depending on the photon emission angle. 
Thus, the conventional TOF counter measures the arrival time 
of the earliest photons 
out of many, whereby the remaining large amount of photons are not 
effectively used for the time measurement. 
While effect (1) cannot be reduced as far as the
scintillation is concerned, effect (2) can be resolved by the 
measuring the arrival times of the earliest photons as a function of the
photon emission angle, since, in the total-reflection-type scintillator 
bar, most photons propagate
in the scintillator while keeping their original emission angles. 
This two-dimensional information could improve the 
time resolution, depending on the number of angular segmentations 
of the photon detector and on the number of photons collected 
at each segment. 

When Cherenkov radiation is utilized, although its number of produced
photons is much smaller than in the scintillation case, effect (1) 
can, in principle, be disregarded. 
In addition, since the Cherenkov photon emission-angle is uniquely
determined by the particle velocity ($\beta$) and since 
the propagation time of photons in 
a light guide of an internal-total-reflection type 
can be calculated as a function of the photon emission angle,  
a measurement of a correlation between TOP and the photon emission anlge 
could provide PID information by itself, as we propose below.

In the DIRC concept [2, 3], 
Cherenkov photons produced by a 
quartz radiator are transported to an end of the radiator by means of 
internal total reflection, and its ring image is magnified and
projected onto a photon-detector plane placed at the bar end. 
A basic study of this concept clearly proved that a photon is 
effectively transported in the long quartz bar, preserving its original 
photon direction. 
The Babar detector at SLAC adopted this concept for PID, and introduced a
large stand-off photon detection system in order to ineffectuate the
finite size of the quartz cross-section onto the ring image [3]. 
On the other hand, Kamae et al. [4] 
proposed a compact
focussing type of DIRC using a small mirror 
instead of a large stand-off system. 
This approach suits particles incident normal to the quartz bar, but
for inclined particles the ring image is distorted due to the finite 
quartz cross-section. 
To make this image-fusing effect insignificant, 
the size and geometry of the focussing mirror become 
unfeasible [5]. 

We propose here a new concept for the Cherenkov-ring image detector, the
TOP detector. 
The two-dimensional information of the ring image is
represented by the TOP of the Cherenkov photon and its horizontal 
angle ($\Phi$; see, Fig.1). 
The use of TOP information with an appropriate focussing mirror
makes the image-fusing effect be disregarded and 
compactification can be realized while maintaining 
the high enough PID ability. 
We describe below the basic concept, characteristic features, practical 
configurations and simulation results on the TOP detector, and 
discuss some issuses for making it a realistic device.

We should mention a paper by Honsheid et al. [6], 
which we 
noticed after our detailed analysis had been completed. 
It discusses the Cherenkov Correlated Timing (CCT) detector: 
CCT [6, 7] 
is a kind of TOF counter with a single 
phototube readout for detecting early arriving photons of 
Cherenkov radiation. 
In ref.[6] 
a similar concept as ours can be found,  
although no detailed study is presented and the focussing mirror 
approach is not introduced. 

\section{Conceptual Design and Expected Properties}
The principal structure of the TOP counter in 
the ($x, y, z$) coordinate system is illustrated in 
Fig.1. 
When a charged particle passes through the radiator bar, Cherenkov
photons are emitted in a conical direction defined by the emission
angle ($\theta_{\rm {C}}$), where $\cos\theta_{\rm {C}}$=1/$n\beta$, 
$n$=refractive index. 
Then, photons propagate to both or either ends by means of
total reflection on the internal bar surface. 
Photons propagating backward are reflected by a flat mirror 
at the end. 
At the forward end, photons are horizontally focused by 
a butterfly-shaped mirror onto a plane where position-sensitive 
multi-anode phototubes are equipped to measure TOP as a function 
of the $\Phi$-angle. 

\subsection{TOP and $\Phi$} 
The TOP is determined only by the z-component of the light velocity 
in the quartz bar as 
\begin{eqnarray}
TOP &=& \left( \frac{L}{c/n(\lambda)} \right)\times
\left(\frac{1}{q_{\rm {z}}}\right),\nonumber\\
 &=& 4.90(ns)\times \frac{L(m)}{q_{\rm {z}}},
\end{eqnarray}
where $L$ is the overall distance from the photon emission point to
the detector, {\it n}$(\lambda)$ is the refractive index of the quartz at 
wavelength $\lambda$, $c$ is the light velocity in a vacuum, and 
$q_{\rm {z}}$ is the directional z-component of the photon emission. 

Denoting the polar and azimuthal angles of an incident charged particle as 
$\theta_{\rm {inc}}$ and $\phi_{\rm {inc}}$ (see, Fig.1), respectively, 
the directional components of the photon velocity 
$q_{\rm {i}}$ ({\rm i}={\it x, y, z}) are written as 
\begin{eqnarray}
q_x &=& q_{x'}\cos\theta_{\rm {inc}}\cos\phi_{\rm {inc}}
	-q_{y'}\sin\phi_{\rm {inc}} 
	 +q_{z'}\sin\theta_{\rm {inc}}\cos\phi_{\rm {inc}}, \nonumber\\
q_y &=& q_{x'}\cos\theta_{\rm {inc}}\sin\phi_{\rm {inc}}
	+q_{y'}\cos\phi_{\rm {inc}} 
	 +q_{z'}\sin\theta_{\rm {inc}}\sin\phi_{\rm {inc}}, \\
q_z &=& -q_{x'}\sin\theta_{\rm {inc}} 
	 +q_{z'}\cos\theta_{\rm {inc}}, \nonumber
\end{eqnarray}
where $q_i$'s ($i$=$x^{\prime}$, $y^{\prime}$, $z^{\prime}$) 
are the directional components of the photon emission in
the frame where the particle moves along the z$^{\prime}$-axis: 
\begin{eqnarray}
q_{x'} &=& \sin\theta_{\rm {C}}\cos\phi_{\rm {C}}, \nonumber\\
q_{y'} &=& \sin\theta_{\rm {C}}\sin\phi_{\rm {C}}, \\
q_{z'} &=& \cos\theta_{\rm {C}}. \nonumber
\end{eqnarray}
The horizontal and vertical photon angles at the bar end are
then given as 
\begin{eqnarray}
\Phi &=& arctan\left(\frac{q_x}{q_z}\right),\\
\Theta &=& arctan\left(\frac{q_y}{q_z}\right). \nonumber 
\end{eqnarray}
The $\phi_{\rm {C}}$ angle of Cherenkov photons distributes uniformly 
over 2$\pi$. When we fix $\phi_{\rm {C}}$, individual three-directional 
components ($q_i$) are uniquely related to $\theta_{\rm {C}}$. 
Therefore, a measurement of any two directional components or 
any two combinations of them among eqs.(3) and (4) provides 
information about $\theta_{\rm {C}}$. 

Accordingly, in a TOP concept, we use the two parameters 
($q_z$, $\Phi$) for 
extracting $\theta_{\rm {C}}$, while in DIRC, 
($\Theta$, $\Phi$) are used. 
The advantage of our approach is that 
the $q_z$ measurement is not seriously influenced by the quartz 
bar cross-section. 

\subsection{Accuracy of the TOP measurement} 
Here, we estimate various contributions to the accuracy of the TOP 
measurement with a detector which we are preparing for R\&D work. 
By modelling the detector as presented in Fig.1, the 
parameters of each element and the expected machining or construction 
accuracy are assumed as described below. 

The size of the radiator bar is 20 mm-thick (in $\it {y}$), 
60 mm-wide (in $\it {x}$), and about 3 150 mm (in $\it {z}$), on which 
a butterfly-shaped 
forcussing mirror block and a flat reflection mirror are 
attached at the forward and backward ends, respectively. 
The focussing mirror comprises two sub-elements, 
each forming an arc of a circle with a radius of 250 mm. 
As a PID detector for the Belle experiment, we suppose these 
detectors to be placed at R=1 m from the beam axis to form 
a cylindrical configuration, and due to an asymmetric-energy 
collider the detector is asymmetrically configured with respect to 
the z-direction to cover the polar angle from 
30$^{\rm {o}}$ to 140$^{\rm {o}}$ in the lab-frame.
For simplicity of calculation, the photon-propagation length 
in the focussing mirror is set as 20 cm for all photons. 

The material of the radiator bar and the mirrors is 
synthetic optical quartz 
(SUPRASIL-P30) manufactured by Shin-etsu Quartz Co. and their 
surfaces are polished by Bikovsky-Japan Co. with a flatness of 
$\pm$5 $\mu$m. 
The quartz has a refractive index of $n$=1.47 at a wavelength 
($\lambda$) of 390 nm and a transmittance of $\>$90\%/m at 
$\lambda$=250 nm. 
The reflecting surface of the mirrors comprises aluminum 
of a few 100 nm-thick on 200 $\rm {\AA}$-thick chromium 
by evaporation, which provides a reflection efficiency of 98\%. 
The photon detector (R5900U-00-L16) is a multi-anode (16 channel) linear 
array phototube with a bialkali window by Hamamatsu Photonics Co. 
Although this tube does not function under a magnetic field, 
it has a quite suitable TTS for our early R\&D work 
without a magnetic field.  \\

Now, by taking the above structure of the TOP counter into account, 
we estimate various contributions to the accuracy of the PID 
measurement. \\

\noindent
{\em (1) $\lambda$-dependence of n}
\par
The refractive index ($n(\lambda)$) of the quartz varies 
depending on $\lambda$. 
As a result, the Cherenkov angle ($\theta_{\rm {C}}$) and the total reflection angle 
vary for individual photons in addition to the propagation velocity. 
The effect to TOP is 
\begin{equation}
\frac{\sigma(TOP)}{(TOP)} = \left(\frac{\sigma_n}{n}\right )
	\times (1-\alpha), 
\end{equation}
where
\begin{equation}
\alpha = n \left(\frac{\partial q_z /\partial n}{q_z}\right).
\end{equation}
Accounting for the $\lambda$-dependence of the Cherenkov photon yield 
and the $\lambda$-dependent quantum efficiency of the phototube, 
we take $n$=1.47$\pm$0.008 ($\Delta n/n$= 5$\times 10^{-3}$) 
for the numerical calculation described in the next subsection. \\

\noindent
{\em (2) Angular resolution} 
\par
The $\Phi$-angle resolution depends on the position resolution 
of the phototube and the focussing accuracy of the mirror. 
The designed focussing mirror has a resolution of 
$\sigma$=0.8 mm with a dispersion of $d\Phi/dx$=0.5$^{\rm {o}}$/mm. 
Since the phototube has a 0.8 mm anode width, as described below, 
the $\Phi$ resolution is evaluated as $\sigma_{\Phi}\approx 0.5^{\rm {o}}$. 

The $\Phi$-dependent behavior of this effect on TOP shows a steep increase 
with $\Phi$, but a very small contribution at small $\Phi$-angles, 
as numerically discussed in the next subsection. 
This limits the $\Phi$ aparture to be $\mid\Phi\mid \leq$ 45$^{\rm {o}}$. \\

\noindent
{\em (3) TTS of the phototube} 
\par
The phototube (R5900U-00-L16) has a 16 mm $\times$ 16 mm photo-sensitive area, 
which is segmented into 16 strips with an area of 0.8 mm $\times$ 
16 mm and a 0.2 mm space between adjacent strips. 
The quantum efficiency ($QE$) is 25\% at 400 nm and 
$\sigma_{\rm {TTS}}\approx$70 ps 
with a pulse rise-time of 0.6 ns 
for a single photoelectron. \\

\noindent
{\em (4) Timing accuracy of the start signal}
\par
We expect a start timing accuracy of 25 ps from the RF signal of 
the accelerator. \\

\noindent
{\em (5) Effect of the quartz thickness} 
\par
Because relativistic particle takes 66 ps to pass through the 20 mm-thick 
quartz, the time spread of the Cherenkov photon emission 
is estimated to be $\sigma_{\rm {thick}}$= 20 ps (=66 /$\sqrt{12}$). 
When the incident particle is inclined with respect to the normal 
of the bar surface, the time-spread becomes much smaller due to 
cancelleration of the particle flight-time and the photon
propagation-time. 
For instance, $\sigma_{\rm {thick}}$ is 6 ps and 10 ps at 
$\theta_{\rm {inc}}$= 
30$^{\rm {o}}$ and 60$^{\rm {o}}$, respectively.\\

\noindent
{\em (6) Effect of the charged-particle tracking accuracy} 
\par
In the Belle detector, the z position of a charged-particle track is
measured by the central drift chamber (CDC) 
with an accuracy of about 2 mm. 
This means that it corresponds to a time uncertainty of 10 ps. 

On the other hand, the accuracy of the momentum measurement is 
related to the accuracy of the Cherenkov angle as 
\begin{equation}
\sigma_{\theta_{\rm {C}}} = \left( \frac{\cot\theta_{\rm {C}}}
{\gamma^2}\right)
\times \left(\frac{\sigma_{\rm {p}}}{p}\right). 
\end{equation}
Since the Cherenkov angle satisfies $\cot\theta_{\rm {C}}\sim 1$ 
for our particles 
and the CDC provides $\sigma_{\rm {p}}$/p$\approx 5\times10^{-3}$, 
$\sigma_{\theta_{\rm {C}}}$ can be ignored compared with 
the difference in the Cherenkov angle between $\pi$ and K.  
The former is on the order of $10^{-2}-10^{-3}$ smaller than 
the latter. 
For instance, $\Delta( \theta_{\rm {C}}(\pi) - \theta_{\rm {C}}(K) )$
= 6.5 mrad, 
while $\sigma_{\theta_{\rm {C}}}$ is $7\times 10^{-3}$ mrad and 
$7\times 10^{-2}$ mrad for 4 GeV/c $\pi$ and K, respectively. 

The accuracy of the incident-angle measurement 
contributes to the accuracy of the TOP measurement as 
\begin{equation}
\frac{\sigma(TOP)}{(TOP)} = -\left(\frac{1}{q_z}\right)\times 
\left(\frac{\partial q_z}{\partial\theta_{\rm {inc}}}\right)
\sigma_{\theta_{\rm {inc}}}
\end{equation}
and 
\begin{equation}
\frac{\sigma(TOP)}{(TOP)} = -\left(\frac{1}{q_z}\right)\times 
\left(\frac{\partial q_z}{\partial\phi_{\rm {inc}}}\right)
\sigma_{\phi_{\rm {inc}}}=0.
\end{equation}
The effect of any $\theta_{\rm {inc}}$ ambiguity on TOP is evaluated 
to be less than 10 ps in the case of the Belle detector [1]. 
It is an order-of-magnitude smaller than the contribution 
from item (3) ($\sigma_{\rm {TTS}}\approx$70 ps), and is thus disregarded. 
Furthermore, it is, on the other hand, a specific feature of our approarch that 
the TOP is independent of $\phi_{\rm {inc}}$, as is obvious from eq.(2). 
Therefore, the $\phi_{\rm {inc}}$ ambiguity does not intrinsically 
affect the TOP measurement. 

\subsection{Expected separability}
From the above, the uncertainty of the TOP measurement arises 
predominantly from the effects of items (1), (2) and (3). 
In order to understand the TOP behavior and to evaluate 
the size of the above effects to TOP and the PID power, 
a calculation is performed with the following conditions. 
The $\lambda$ dependence of the refractive index is disregarded, 
but its effect is represented by $n$$=1.47 \pm 0.008$ in eq.(5); 
the $\Phi$ resolution is set as $\sigma_{\Phi}=0.5^{\rm {o}}$; 
the effects of the items (4) and (5) and the digitization bin of 25 ps 
are included into (3) to make 
an effective $\sigma_{\rm {TSS}}$=80 ps 
(denoted hereafter as (3)$^{\prime}$); 
the efficiencies of photon propagation and reflection are assumed 
to be 100\%. 

For PID in the collider geometry, in addition to the TOP measurement, 
we added to the TOP the particle's TOF from the colliding point 
to the quartz radiator placed at R=1 m. \\

\noindent
{\em 2.3.1.~~$\delta$(TOP+TOF)
and individual effects} 
{\ }
\par
In Fig.2, the individual uncertainties of items 
(1), (2) and (3)$^{\prime}$ are shown for various cases of 
the incident angle ($\theta_{\rm {inc}}$), 
in addition to the TOP+TOF and TOP differences between 4 GeV/c 
$\pi$ and K (denoted hereafter as $\delta$(TOP +TOF)
and $\delta$(TOP), respectively).  
(The propagation length ($L$) in Figures (a-e) is uniquely determined 
for each $\theta_{\rm {inc}}$ by assuming the particle 
coming from the colliding point under our geometory. 
On the other hand, the length ($L$) in Figures (f, g) is not real, but 
hypothetical, only for discussion purposes.) 
The asymmetries seen in the figures are caused by deflection 
of the azimuthal angle 
of the incident particle due to an axial magnetic field of 1.5 T. 

At normal incidence (see, Fig.2(c), (d) and (g)), 
effect (1) is an order of magnitude smaller than $\delta$(TOP); 
this relation holds irrespective of $L$, 
because of their proportionality to $L$. 
Effect (2) has a strong $\Phi$ dependence.  
In the region of the small $\Phi$-angle the effect is negligible, 
while in the large-$\Phi$ region it becomes dominant in determining 
the resolution. 
Mostly from this fact, the $\Phi$-angle aperture is limitted 
to $\mid\Phi\mid \leq 45^{\rm {o}}$, where the TOP measurement uncertainty 
is comparable to or less than $\delta$(TOP+TOF) under most of 
the incident-particle conditions. 
On the other hand, because $\theta_{\rm {inc}}$ deviates from normal incidence 
the value of $\alpha$ decreases, and item (1) gives almost 
the same order of magnitude as $\delta$(TOP+TOF)
at $\theta_{\rm {inc}}=30^{\rm {o}}$ (Fig.2(a)), and several to 
10-times larger size 
at $\theta_{\rm {inc}}=140^{\rm {o}}$ (Fig.2(e)). 

Note that the sign of $\delta$(TOP) at small $\Phi$ reverses at 
around $\theta_{\rm {inc}}\sim 45^{\rm {o}}$ or $\theta_{\rm {inc}}
$-90$^{\rm {o}}\sim$ 45$^{\rm {o}}$, 
since $\theta_{\rm {inc}}$ almost 
coincides with the Cherenkov angle ($\sim 45^{\rm {o}}$) and then 
the relative magnitude of $q$$_z$ of Cherenkov photons from $\pi$ and K 
reverses (see, Fig.2(a, e, f)). 
Thus, in the case of large $L$, as in Fig.2(f), 
$\delta$(TOP) is cancelled with $\delta$(TOF), which reduces 
the PID power compared with the small $L$ case, as in Fig.2(a). \\

\noindent
{\em 2.3.2.~~$\pi$/K separability}
{\ }
\par
In the previous discussion, the expected resolution for the TOP 
measurement is given for the detection of a single Cherenkov photon. 
However, when we use the position-sensitive photon detector, we can 
use every Cherenkov photon to obtain PID information. 
In order to estimate the PID performance, we define 
the $\pi$/K separability ($S$$_{\rm {o}}$) as 
\begin{equation}
S_{\rm {o}} = \sqrt{\left \{ \sum_{\rm {i}}\left(\frac{\delta(TOP+TOF)
^i}   {(\sigma_{\rm {T}})^{\rm {i}}}\right)^2\right \}\cdot \kappa}. 
\end{equation}
In the calculation, a single Cherenkov photon is assigned at every 
one-degree in $\phi_{\rm {C}}$ over a range of $2\pi$, and individual 
$\delta$(TOP+TOF)$^i$ and the total uncertainty $(\sigma_{\rm {T}})^i$ 
arising from the above-mentioned items are obtained 
for every $i$-th photon. 
The summation is taken over for the photons with 
$\mid\Phi\mid\leq$45$^{\rm {o}}$, which are totally reflected on 
the internal surface of the quartz bar. 
Because the number of Cherenkov photons for the 20 mm-thick quartz bar 
is fixed as $N_{\rm {o}}^{\gamma}=160/(\sin\theta_{\rm {inc}}\times 
QE)$ with $QE$=25\%, 
$\kappa$ is a normalization factor of $N_{\rm {o}}^{\gamma}/360$. 

In Fig.3 the resultant $S$$_{\rm {o}}$ for $\pi$/K is presented as 
a function of the particle momentum ($p$) for three different 
$\theta_{\rm {inc}}$'s. 
For a normal incident particle, two cases are considered separately: 
(a) Cherenkov photons directly come to the photon detector (FW), 
or (b) go to the backward end and then return to the detector (BW). 
A better separation is obtained for (b) because of 
the longer propagation distance of photons. 
The incident-angle dependence of $S$$_{\rm {o}}$ is shown in Fig.4 for $p$=2, 4 and 
5 GeV/c particles, where the FW and BW cases are 
indicated by the solid and dotted lines, respectively. 
The decrease in $S_{\rm {o}}$ around $\theta_{\rm {inc}}$=50$^{\rm {o}}$ is 
due to destructive correlation between $\theta_{\rm {inc}}$ and 
$\theta_{\rm {C}}$, as mentioned above. 
The $S_{\rm {o}}$ variation on the propagation length ($L$) is calculated 
by varing the quartz length for $\theta_{\rm {inc}}$= 30$^{\rm {o}}$, 
60$^{\rm {o}}$ 
and 90$^{\rm {o}}$(FW), as shown in Fig.5.  
At normal incidence, $\delta$(TOP+TOF) becomes large 
relative to the effects of (2) and (3)$^{\prime}$ as $L$ increases, 
and, accordingly, $S_{\rm {o}}$ is improved. 
On the other hand, at around $\theta_{\rm {inc}}$=30$^{\rm {o}}$ 
the effect (1) becomes large with $L$, and $S_{\rm {o}}$
decreases, as can be seen in Fig.2(a) and (f). 

\section{Simulation Study}
Based on the basic concepts considered in the preceeding 
section, we carried out a simulation study on the realistic
environment. 
Photons are generated following to the Cherenkov spectrum 
(d$N$$(\lambda)$/d$\lambda$) convoluted by the quantum 
efficiency $QE$($\lambda$) of the phototube. 
The Cherenkov angle is determined by using the $\lambda$-dependent 
quartz refractivity ($n$($\lambda$)). 
The effect of the quartz thickness is naturally 
implemented in the simulation procedure. 
The effective TTS of the phototube is therefore set 
to be $\sigma_{\rm {TTS}}=$75 ps, instead of the previous value of 80 ps, 
including the ambiguity of only the timing reference signal; 
this time resolution is considered in 
smearing the calculated TOP+TOF values with a Gaussian of 
$\sigma=\sigma_{\rm {TTS}}$. 
The phototube is assumed to detect only the earliest photon 
if multi photons hit the same anode channel. 
The focussing resolution is also included by smearing the calculated 
$\Phi$-angle with a Gaussian of $\sigma_{\Phi}=0.5^{\rm {o}}$ as well. 
Digitizations of the arrival time and $\Phi$-angle of photons are 
made in 25 ps and 0.5$^{\rm {o}}$ bins, respectively. 
All other treatments are the same as in the previous calculation. 

To display the various results for $\pi$ and K tracks with 
the same momentum, we set parameters (as an example, $p$=4 GeV/c and 
$\theta_{\rm {inc}}=90^{\rm {o}}$) and 
require to detect only FW photons, unless the conditions are specified. 
Figure 6(a) shows a typical $\Phi$-angle distribution 
of the number of detected photons for the $\pi$ track. 
The number of produced photons is 137 in total, 
while the detected FW photons within $\mid\Phi\mid$=45$^{\rm {o}}$ are 33. 
The photons distribute rather uniformly with an average of less 
than one per bin ($\Delta\Phi$=0.5$^{\rm {o}}$). 
The TOP+TOF is shown in Fig.6(b), where the circles and crosses are 
the detected photons radiated by the $\pi$ and K track, respectively. 
They form two slightly aymmetric parabola-like distributions due to 
an inclined $\phi_{\rm {inc}}$-angle by a deflection effect due to 
the magnetic field. 
It is 4.3$^{\rm {o}}$ in this case. 
The dotted curves are the TOP+TOF calculated in the previous 
section by disregarding the $\lambda$ dependences and 
generating photons with $\theta_{\rm {C}}$=1/$n$$\beta$ ($n$=1.47). 
The upper two dotted curves in the figure correspond to 
the calculated TOP+TOF for the BW photons. 

The deviations of TOP+TOF for an incident $\pi$ and K from the calculated 
distribution assuming $\pi$, $\Delta$(TOP+TOF)$_{\pi/K}(\pi)$, are 
presented in Fig.6(c). 
In finding the deviation, two solutions exist at every $\Phi$-angle 
due to the above-mentioned deflection effect, 
so that we choose the smaller deviation as the correct one. 
The circles and closes are for photons emitted by a $\pi$ and 
a K, respectively. The former distributes around the null value, 
while the latter shifts to the positive side, as expected. 

The solid and dotted histograms in Fig.7 show the distribution of 
deviations, $\Delta$(TOP+TOF)$_{\rm {\alpha}}(\beta)$, for photons 
within $\mid\Phi\mid\leq$45$^{\rm {o}}$, where $\alpha$, 
$\beta$=$\pi$ or K; also, $\alpha$ is the particle species 
of the track and 
$\beta$ is the particle species assumed. 
The solid histogram presents the results under the correct assumption and 
the dotted histogram presents those under an incorrect assignment. 
Figures 7(a) and (b) show typical deviations for 
an event, and Figs.7(c) and (d) present the results for the accumulation 
of 100 tracks.  
We obtain the probability function ($P$$(\pi$/K)) 
by fitting the results of the right assumption in Figs.7(c) and (d) 
with Gaussian functions, as shown in the figures by the solid curves. 

Based on this probability function, a log-likelihood ($ln\cal L$) 
under the assumption of $\pi$ and K's are calculated for 400 tracks 
each for incoming $\pi$ and K. 
Figure 8(a) shows a scatter plot of the log-likelihoods thus obtained 
in the case of $p$=4 GeV/c and $\theta_{\rm {inc}}$= 90$^{\rm {o}}$ 
for the FW photons;  
a clear separation between $\pi$ and K can be seen. 
The separability ($S$) is calculated by taking the difference in 
the log-likelihood ($\Delta ln\cal L$) between the two assumptions 
(see Fig.8(b)) and using $S$=$\sqrt{2\Delta ln\cal L}$; 
$S$=5.7 in this case.
Thus, the calculated $S$'s for different $p$'s and 
$\theta_{\rm {inc}}$'s are 
indicated in Fig.9 for both FW and BW photons. 

\section{Discussion and Summary}

We have shown that a high PID capability is attainable by using 
a (TOP, $\Phi$) measurement of the Cherenkov ring image. 
Although R\&D is necessary, especially, 
on a high-quality position-sensitive phototube operable 
in a magnetic field and on an optimum focussing mirror, 
a good $\pi$/K separation is well 
expected at the Belle barrel detector.  

Some remarks are given below: 
\begin{itemize}
\item
The development of a position-sensitive phototube is most 
important to realize the separability so far presented. 
It has twofold issues. 
One is to make it operable under a magnetic field and 
we have being developed a fine-mesh-type multi-anode phototube 
for this purpose. 
The other is to remove or reduce the dead space dominated 
by the phototube package. 
The phototube (R5900U-00-L16) has a surface size of 30 mm$\times$30 mm 
in which the sensitive area is 16 mm$\times$16 mm, so that 
the effective area at our R\&D counter becomes only about 40\%. 
About half is in horizontal and 80\% in the vertical. 
The required size of the phototube is 20 mm$\times$100 mm with 
a bin step of 1 mm for detecting 
photons with $\mid\Phi\mid\leq$45$^{\rm {o}}$. \\

\item
Adding the TOF information to the TOP information helps 
the PID, especially for low-momentum particles. 
However, for high-momentum particles its effect is not appreciable 
in our configuration. 
For instance, at 4 GeV/c tracks with $\theta_{\rm {inc}}$=90$^{\rm {o}}$, 
$\delta$(TOF) are 23 ps, which is only about 1/3 of $\sigma_{\rm {TTS}}$. 
However, by setting a longer flight-length (R), a sizable improvement 
can be realized, as listed in Table 1; in such cases the TOP 
detector acts as a kind of a high-resolution TOF counter by means of 
Cherenkov radiation, since the Cherenkov-angle difference does not 
produce sufficient $\delta$(TOP). \\

\item
The $\pi$/K separability by the TOP measurement has 
an angular dependence, as can be seen in Fig.9, while DIRC gives 
a more-or-less flat distribution with the incident angle. 
Especially, TOP shows a separability decrease at 
$\theta_{\rm {inc}}\sim$ 50$^{\rm {o}}$, and it is in principle inevitable. 
On the other hand, TOP can give a larger $S$ at normal incidence. 
Furthermore, in this case, TOP has the merit of doubling $S$ by 
detecting the both FW and BW going photons those of which have 
an arrival time difference of about 20 ns, so that they can be easily 
identified by using a multi-hit TDC with a high resolution. \\

\item
One possible way to improve the separability for tracks with 
a large incident angle is to equip an inclined surface, say, 
45$^{\rm {o}}$ to the reflection mirror at the backward end. 
This will make the reclining Cherenkov ring image stand up to 
the normal direction, so that it becomes a case close to a normal 
incidence track and photons propagate the full bar length. 
Although an opposite effect occurs for a track with 
$\theta_{\rm {inc}}\sim$90$^{\rm {o}}$, 
a rather sufficient TOP difference is produced by propagating 
a certain long distance before the photons arrive at 
the reflection mirror.  
Optimization of the inclining angle of the reflection mirror 
should be examined. \\

\item
Because the present design of the focussing mirror has 
a horizontally-extended structure, it is not suitable as 
a barrel detector, and therefore 
we need to find a practical optical structure for the compact 
system. 
Although the vertically configured mirror avoids this geometrical 
problem, the choise of the horizontal configuration is due to the fact 
that the TOP+TOF distribution for the vertical configuration 
has a larger differentiation to the angle $\Theta$, as shown in Fig.10, 
so that the TOP+TOF measurement uncertainty becomes larger for an angle 
resolution of $\Delta\Theta$ at the same accuracy as $\Delta\Phi$. 
\par
The simplest and practical solution for the above is to 
tilt the counter around the z-axis by an angle of, for instance, 
18.4$^{\rm {o}}$(=arctan(2 cm-thick/6 cm-width)) in our case, so that 
every mirror can avoid to interfere with each other. 

\par
A TOP counter with a butterfly-shaped mirror, illustrated in Fig.1, 
has a symmterical geometry with respect to the z-axis. 
By dividing the butterfly mirror along the z-axis, 
a half butterfly-shaped counter can be 
constructable, although the bar width also becomes half so as to have 
sufficient $\Phi$-angle resolution.  \\

\item
For simplicity of the calculation, the probability function 
(P$(\pi$/K)) is obtained in the previous section from the whole 
$\Delta$(TOP+TOF) distributions within $\mid\Phi\mid$=45$^{\rm {o}}$, 
while the spread of the distribution has the $\Phi$-dependence. 
Therefore, if we properly take account of this $\Phi$-dependence in the 
probability examination, better separability will be obtained. 
For instance, the rms of the distribution varies from 85 ps 
to 140 ps within the accepted $\Phi$-angle region in the case of 
4 GeV/c normal incident tracks. \\

\item
We assumed the detection of only the earliest single photon at 
each phototube anode in the previous section. 
It causes the loss of some photon fraction. For instance, it is 
about 30\% of $\approx$ 110 detectable photons at 
$\theta_{\rm {inc}}$=30$^{\rm {o}}$, 
while it is about 10\% of $\approx$56 and 
$\approx$30 photons at $\theta_{\rm {inc}}$=60$^{\rm {o}}$ 
and 90$^{\rm {o}}$, 
respectively. \\

\item
It may be interesting to use a scintillator instead of the quartz 
material, as mentioned at the beginning, although the situation is 
somewhat obscure without a detailed study. 
Because of its finite decay-time and 
non-unique photon emission-angle, the arrival-time distribution 
at individual $\Phi$-angles will spread. 
However, it is the usual situation at the ordinary TOF counter. 
Since a large number of photons are provided in this case, 
when an optimized configuration is found to have a sufficient number 
of photons at each $\Phi/\Theta$-segmentation, the 
TOF resolution could be improved by a factor of square-root of 
the number of detection segments. 
Angular segmentation could be arranged wider than our TOP case, so that 
the focussing mirror could be smaller and the number of phototubes 
could be less. 
Needless to say, the surface of the scintillator 
should be polished enough to allow photons to be internally reflected. 
\end{itemize}

Although the above-mentioned issues concerning the phototube and 
the mirror structure remain, the proposed detector can be fully 
applicable as a PID device for experiments under a non-magnetic 
enviornment, mostly for fixed-target experiments. 
At fixed-target experiments, because incoming particles are incident 
nearly normal to the detector, the maximum separation 
power can be extracted. 

With the use of a horizontal focussing mirror together with the fact 
that the time ambiguity of Cherenkov radiation along the pass length 
is disregarded compared with the phototube TTS, 
the cross section of the quartz radiator does not have any effects on TOP 
so that we can successfully avoid the unwanted 
image-fusing effect. 
 
The DIRC counter [2, 3, 4] 
verified that 
the Cherenkov image is well-transported by internal total reflection, 
and the CCT counter [6, 7] 
exhibited that a time measurement 
for such Cherenkov photons is feasible. 
While the TOP detecter proposed here is based on a comprehensive 
compilation of the technical verification by those detecters, 
it can be used to exploit a new approach for measuring the 
TOP and $\Phi$-angle correlation by making use of the 
horizontal focussing mirror. 
As a result, it would not only provide high particle identification 
ability, but also a more compact and flexible detector compared 
with the DIRC and the gas Cherenkov imaging detecter.

{\ }\\
{\ }\\

\noindent
{\bf Acknowledgements}
{\ }
\par
We greatly thank Prof. Yoshitaka Kuno for informing us about 
the paper on the CCT counter. 
We also thank Prof. Katsumi Tanimura for his consultation on optics 
relating to our materials and devices. 
This work was supported by Grant-in-Aid for Scientific Research on 
Priority Areas (Physics of CP violation) from the Ministry of
Education, Science, and Culture of Japan. 
{\ }\\

\newpage

\vspace*{5cm}
\begin{table}[h]
\begin{center}
\caption{Achievable separabilities (S) with various flight-lengths (R). 
The simulation results on our TOP detector, separately detecting FW and BW going 
photons, are listed for tracks with $\theta_{\rm {inc}}$=90$^{\rm
{o}}$ 
under no magnetic field. 
The differences of $\theta_{\rm {C}}$ between $\pi$ and K are 6.5 and 1.0 mrad 
for p=4 and 10 GeV/c, respectively, and those of the TOF are 231 and
37 ps at R=10 m.}
\vspace*{3mm}
\begin{tabular}{c|rrrr}\hline
{Flight length R(m)} &\multicolumn{4}{c}{Separability S}\\  
  \cline{2-5}
 &\multicolumn{2}{c}{p=4 GeV/c} &\multicolumn{2}{c}{p=10 GeV/c} \\
 &~~~~FW &~~~~BW &~~~~FW &~~~~BW \\ \hline
0	&5.7	&8.3	&1.5	&1.5 \\
1	&7.4	&9.0	&1.5	&1.5 \\
5	&11.7	&11.6	&1.9	&2.2 \\
10	&17.6	&14.5	&3.0	&2.9 \\
\hline
\end{tabular}
\end{center}
\end{table}
\newpage
{\bf Figure Captions} \\

\begin{description}
\item{Fig.1: } (a) Schematic structure of the TOP detector. 
	The detector comprises a quartz radiator-bar 
	(20 mm-thick, 60 mm-wide, 3150 mm-long) and two mirrors for 
	focussing (the butterfly-shaped one at the forward end) and reflecting 
	(the flat one at the backward end). The Cherenkov photons are 
	horizontally focussed onto the photon-detector plane, and their time 
	and angle information (TOP+TOF and $\Phi$) are measured by
	position-sensitive multi-channel phototubes. 
	Figure (b) indicates the relation among the basic
	variables in the (x, y, z) coordinate system. 
	$\theta_{\rm {C}}$ is the Cherenkov angle. 
	OP indicates the particle trajectory, the direction of which 
	is defined by
	$\theta_{\rm {inc}}$ and $\phi_{\rm {inc}}$. 
	OQ is the unit velocity vector of a Cherenkov photon, 
	of which the direction is defined by $\Phi$ and $\Theta$ and 
	its directional components are denoted as $q_x$, $q_y$ and $q_z$. \\

\item{Fig.2: } TOP+TOF and TOP differences between 
	$\pi$ and K, $\delta$(TOP+TOF) and $\delta$(TOP), 
	with a momentum of 4 GeV/c and expected time uncertainties 
	from effects (1), 
	(2) and (3)$^{\prime}$ for 7 different conditions. 
	Individual conditions are indicated at the top of each figure. 
	Especially, (a)-(e) correspond to the condition that the particle produced 
	at the colliding point with $\theta_{\rm {inc}}$ flies R=1 m under
	a 1.5 T magnetic field and hit the TOP detector with the
	geometry illustrated in Fig.1. 
	The Cherenkov photon which propagates directly to the 
	forward(FW)/backward(BW) ends are considered to be detected 
	separately. 
	The negative value of the $\delta$(TOP) is plotted by 
	the dotted curves in (a), (e) and (f). 
	Details are described in the text. \\

\item{Fig.3: } Expected $\pi$/K separability ($S_{\rm {o}}$) 
	vs the momentum ($p$) at 
	different incident angles ($\theta_{\rm {inc}}$). 
	The solid and dotted curves correspond to the detection of 
	forward(FW) and backward(BW)-going photons, respectively. 
	The approximate numbers of detected photons are: 
	130 at $\theta_{\rm {inc}}$=30$^{\rm {o}}$, 
	39 at $\theta_{\rm {inc}}$=90$^{\rm {o}}$ 
	for both FW and BW photons, and 76 at 
	$\theta_{\rm {inc}}$=140$^{\rm {o}}$ 
	for $p$=4 GeV/c particles. \\

\item{Fig.4: } Expected $\pi$/K separability ($S_{\rm {o}}$) as a function of 
	$\theta_{\rm {inc}}$ for $p$=2, 4 and 5 GeV/c. 
	In the regions of $\theta_{\rm {inc}}\leq$90$^{\rm {o}}$ 
	and $\geq$90$^{\rm {o}}$, 
	only the FW and BW-going photons are considered, respectively.  \\

\item{Fig.5: } Expected $\pi$/K separability ($S_{\rm {o}}$) as a function of 
	the propagation length ($L$) for 
	$\theta_{\rm {inc}}$=30$^{\rm {o}}$, 
	60$^{\rm {o}}$ and 
	90$^{\rm {o}}$ and $p$=4 GeV/c. 
	 L is varied and only FW-photons are considered. 
	The behavior of $S_{\rm {o}}$ is discussed in the text. \\

\item{Fig.6: } Simulated results for a $\pi$ and K track with 
	$p$=4 GeV/c and $\theta_{\rm {inc}}$=90$^{\rm {o}}$. 
	Only FW-going photons are considered. 
	(a) Detected number of photons vs $\Phi$ for a $\pi$ track. 
	In this case total number of photons produced is 137 
	and those detected 
	within $\mid\Phi\mid\leq$45$^{\rm {o}}$ is 33. 
	(b) TOP+TOF as a function of $\Phi$ for $\pi$ ($\circ$) 
	and K ($\times$). The dotted curves indicate the calculated 
	TOP+TOF, of which details are described in the text. 
	(c) Deviation of TOP+TOF from the calculated value assuming the 
	pion, $\Delta$(TOP+TOF)$_{\pi/ K}(\pi)$, 
	for $\pi$ ($\circ$) and K ($\times$).\\

\item{Fig.7: } $\Delta$(TOP+TOF)$_{\alpha}(\beta)$ distributions with 
	$\mid\Phi\mid\leq$45$^{\rm {o}}$ for FW-going photons for tracks 
	of $p$=4 GeV/c and $\theta_{\rm {inc}}$=90$^{\rm {o}}$. 
	Solid (dotted) histogram presents the results under the
	correct (incorrect) assumption of $\alpha=\beta$ 
	($\alpha\neq\beta$). 
	Tracks are assumed as $\beta=\pi$ in (a) and (c), 
	and as $\beta$=K in (b) and (d) to calculate the expected TOP+TOF. 
	(a) and (b) are for the single tracks, while (c) and (d) are 
	obtained by accumulating 100 tracks. 
	Distributions with the right hypothesis are used as 
	the probability function P($\pi$/K), whose functional form 
	is obtained by fitting with Gaussian functions, 
	as shown by the solid and dotted curves in (c) and (d), respectively. \\

\item{Fig.8: } Log-likelihood distributions of detected FW photons with 
	$\mid\Phi\mid\leq$45$^{\rm {o}}$ for 
	$\pi$ and K tracks with $p$=4 GeV/c and 
	$\theta_{\rm {inc}}$=90$^{\rm {o}}$. 
	(a) Two-dimensional log-likelihood distributions for each 
	400 $\pi$ ($\circ$) and K ($\times$) tracks are plotted 
	for hypothesizing $\pi$ (horizontal axis) and K (vertical axis). 
	The dotted line indicates the equal $ln \cal {L}(\pi)$=$ln \cal {L}$(K) line. 
	(b) Distributions of log-likelihood difference are shown 
	for each 400 $\pi$ (solid histogram) and K (dotted histogram) 
	tracks. Curves are the Gaussian fitted to
	the data. The resulting $S$ is 5.7 for this case. \\

\item{Fig.9: } Achievable separability ($S$) vs $\theta_{\rm {inc}}$ for 
	$p$=2, 3, 4 and 5 GeV/c $\pi$ and K tracks. 
	For the regions of $\theta_{\rm {inc}}\leq$90$^{\rm {o}}$ 
	and $\geq$90$^{\rm {o}}$, the FW (solid curves) and BW 
	(dotted curves)-going photons are detected, respectively.  
	The $S$ indicated by the plusses at 
	$\theta_{\rm {inc}}$=90$^{\rm {o}}$ are by detecting both FW 
	and BW photons: $p$=2, 3, 4 and 5 GeV/c from the upper to 
	the lower point.  \\

\item{Fig.10: } TOP+TOF vs $\Theta$. 
	TOP+TOF of the same photons in Fig.6(b) is plotted 
	with $\Theta$ instead of the $\Phi$-angle. 
	The dotted curves indicate the calculated 
	TOP+TOF, whose details are described in the text. 


\newpage

\begin{figure}
\vspace{24cm}
\hspace{-3cm}
\includegraphics{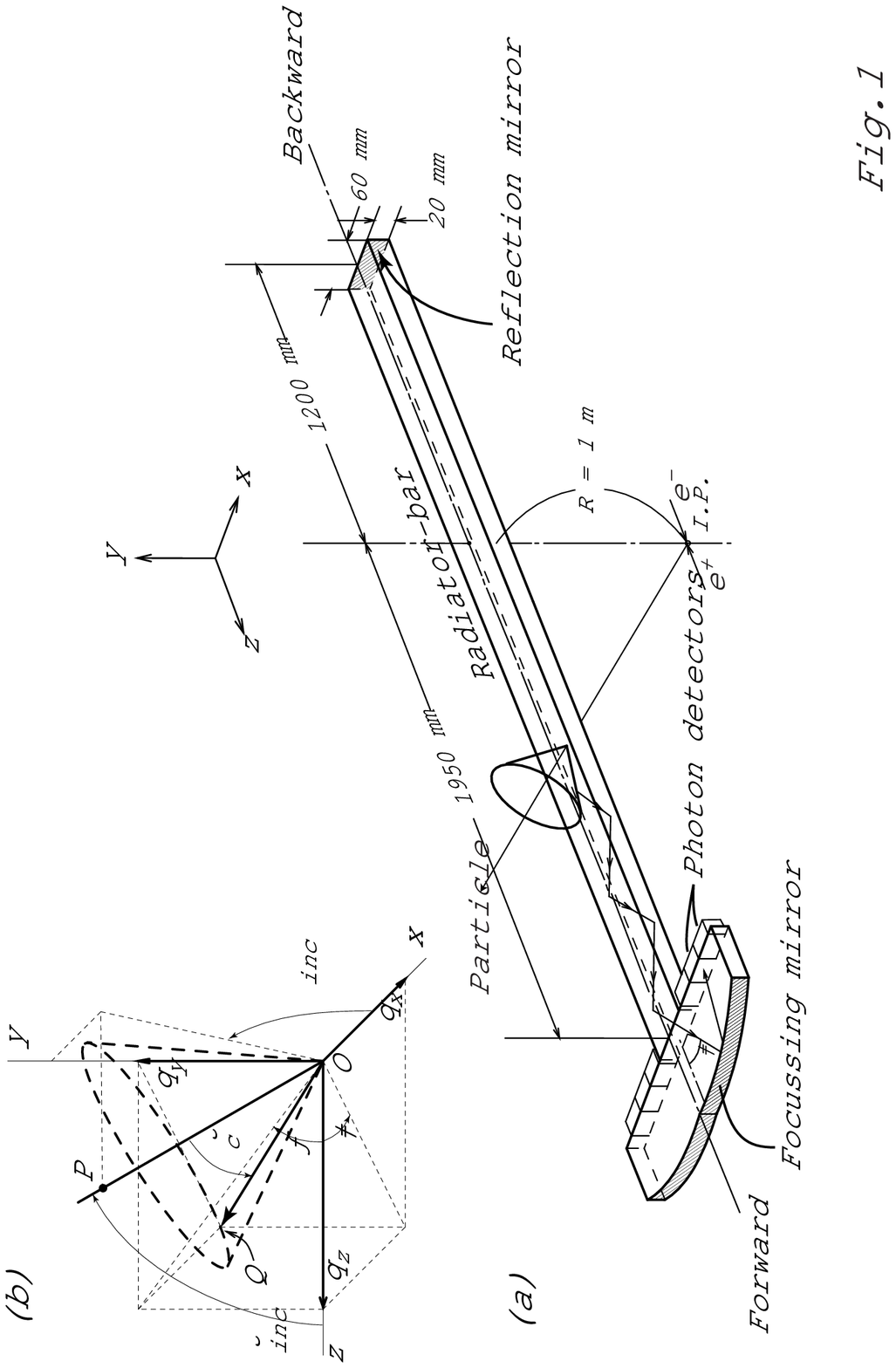}
\end{figure}

\newpage

\begin{figure}
\vspace{24cm}
\hspace{-3cm}
\includegraphics{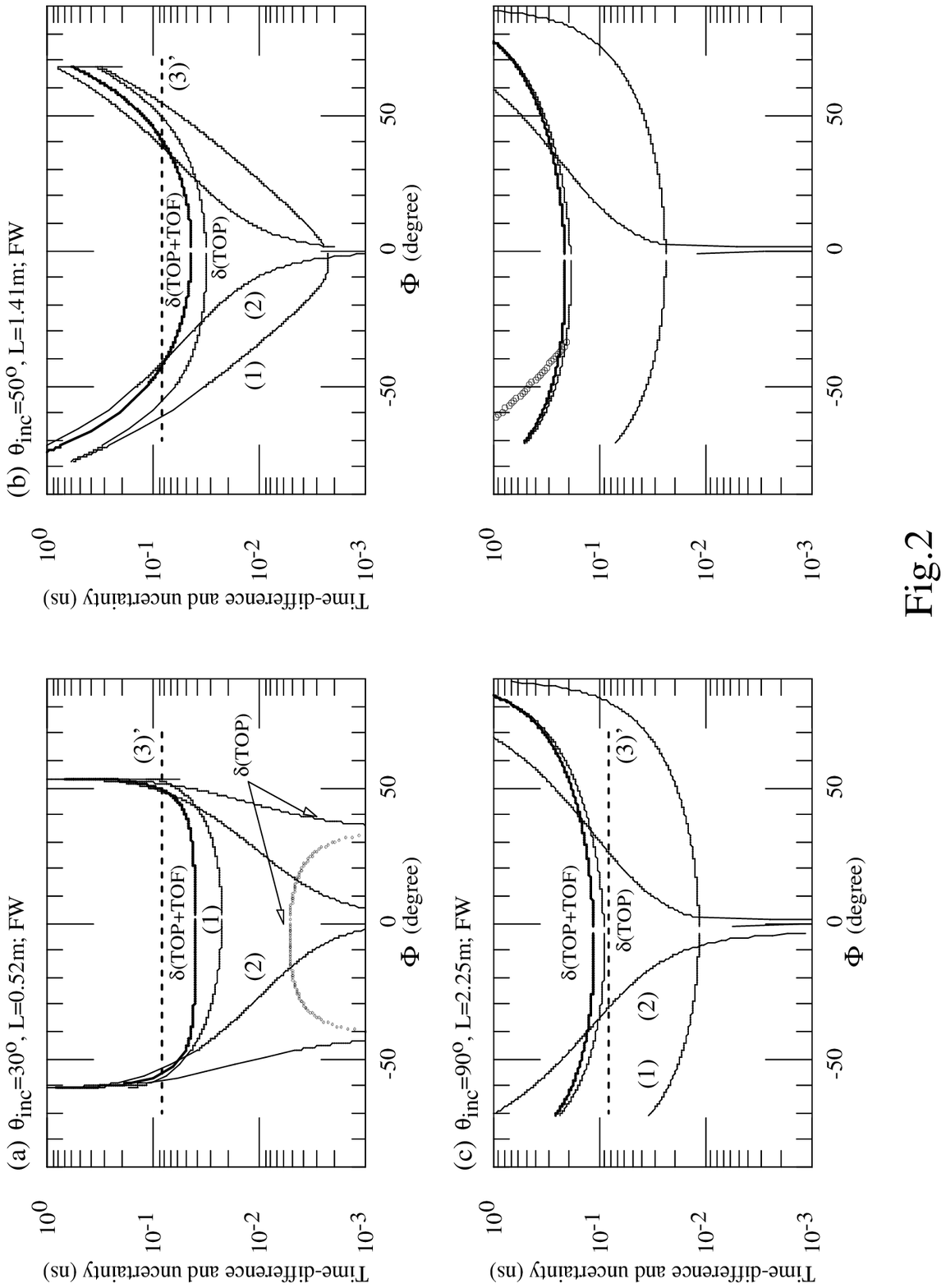}
\end{figure}

\newpage

\begin{figure}
\vspace{24cm}
\hspace{-3cm}
\includegraphics{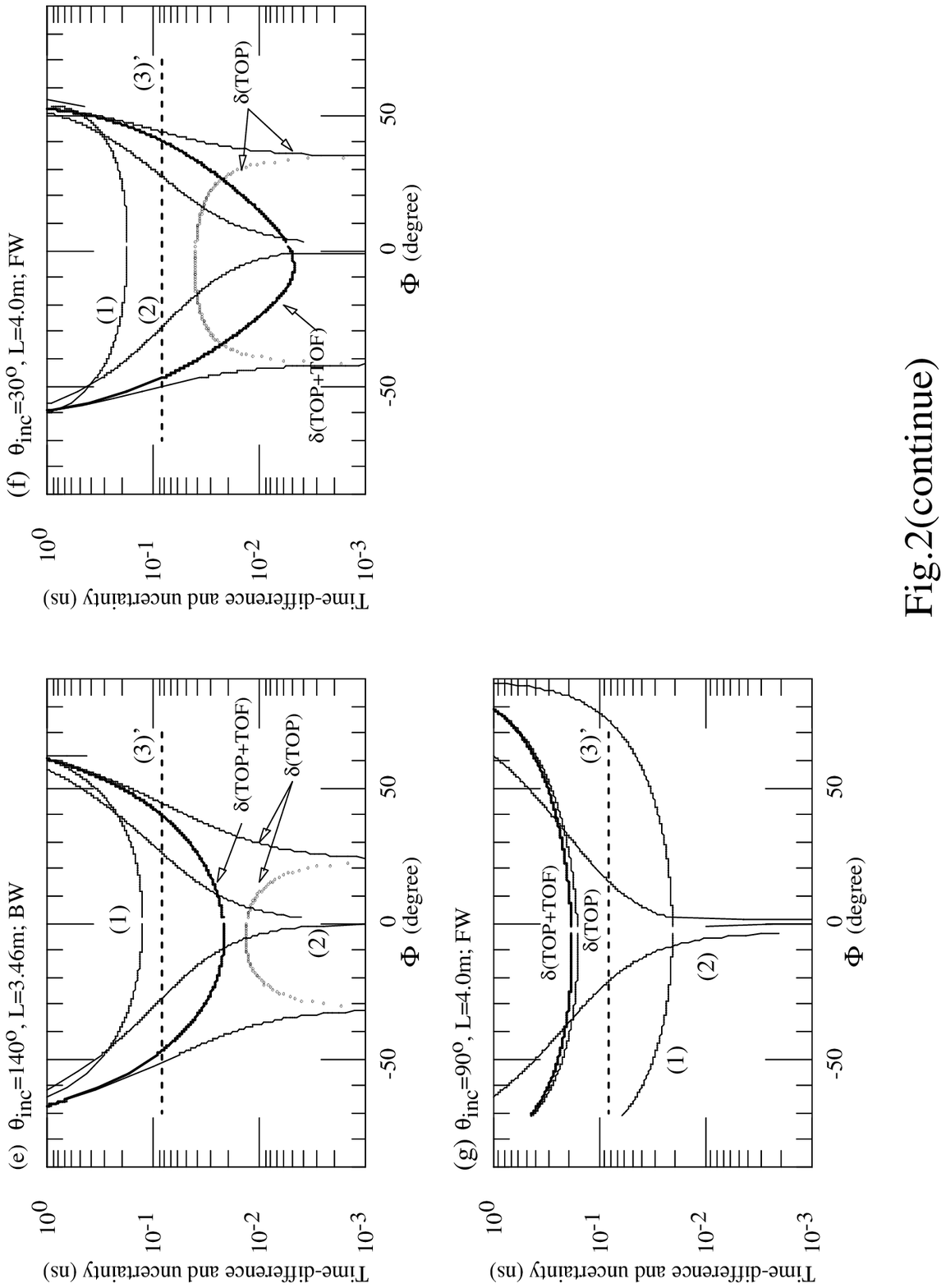}
\end{figure}

\newpage

\begin{figure}
\vspace{24cm}
\hspace{-2cm}
\includegraphics{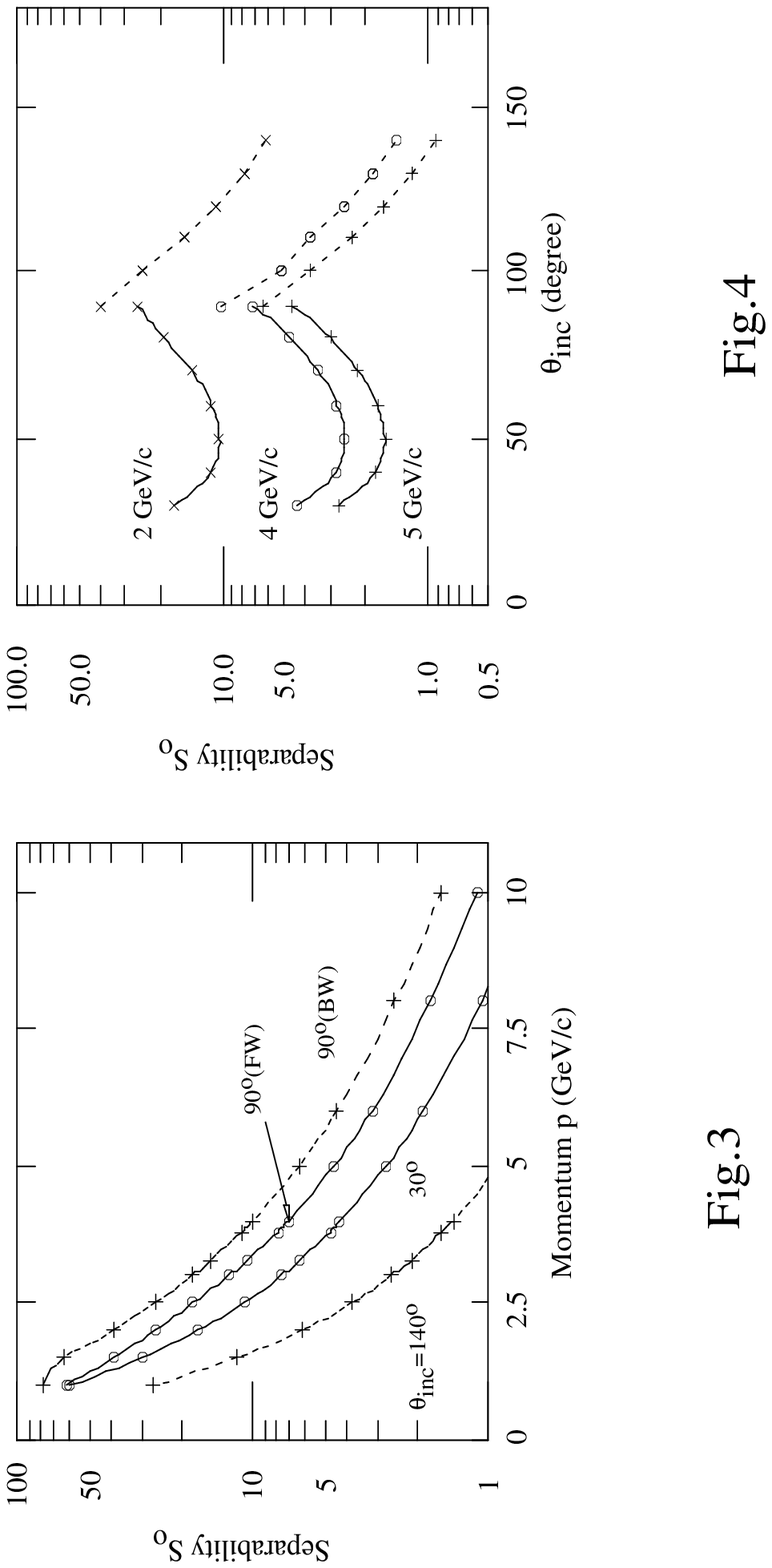}
\end{figure}

\newpage

\begin{figure}
\vspace{24cm}
\hspace{-3cm}
\includegraphics{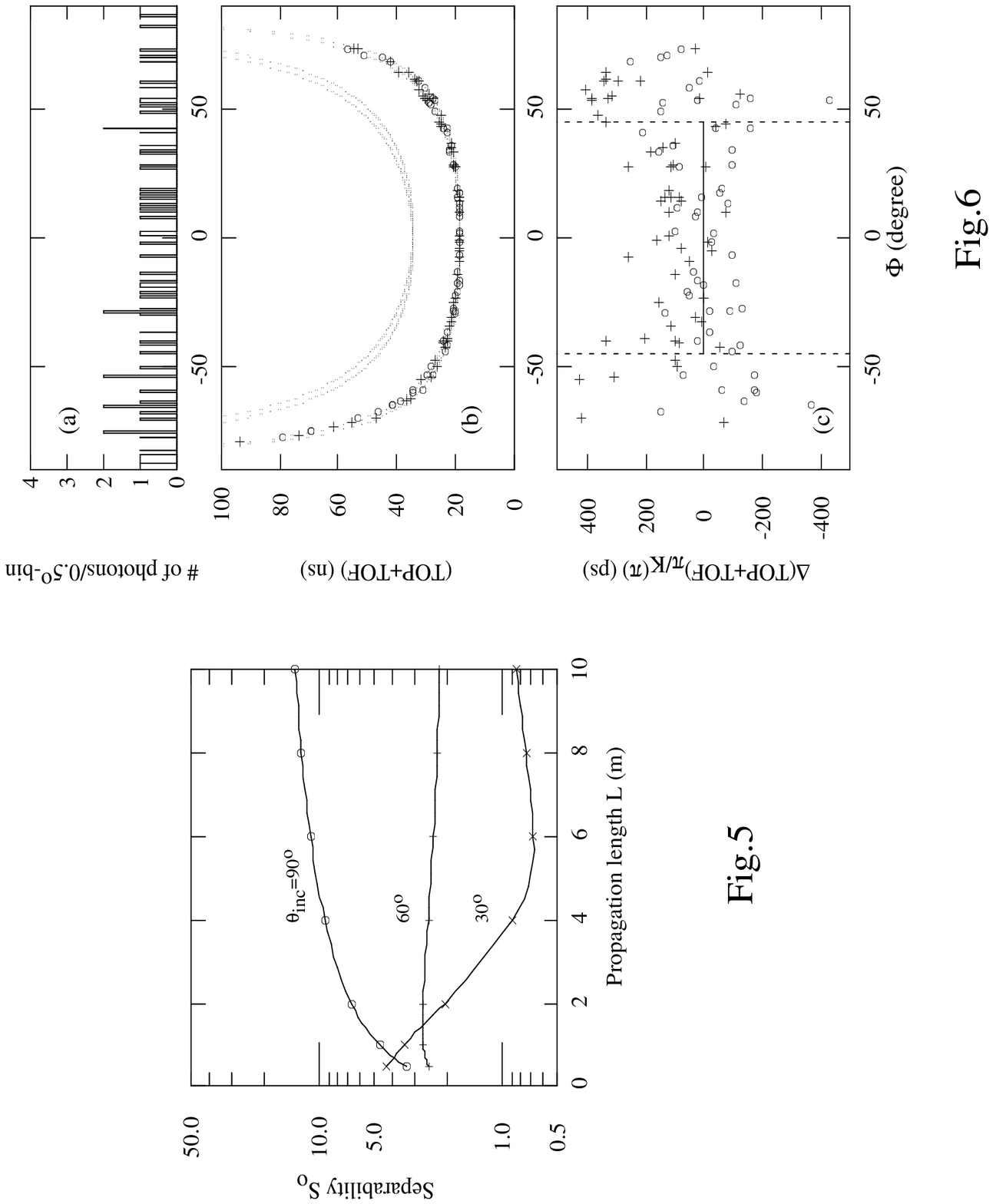}
\end{figure}

\newpage

\begin{figure}
\vspace{24cm}
\hspace{-2cm}
\includegraphics{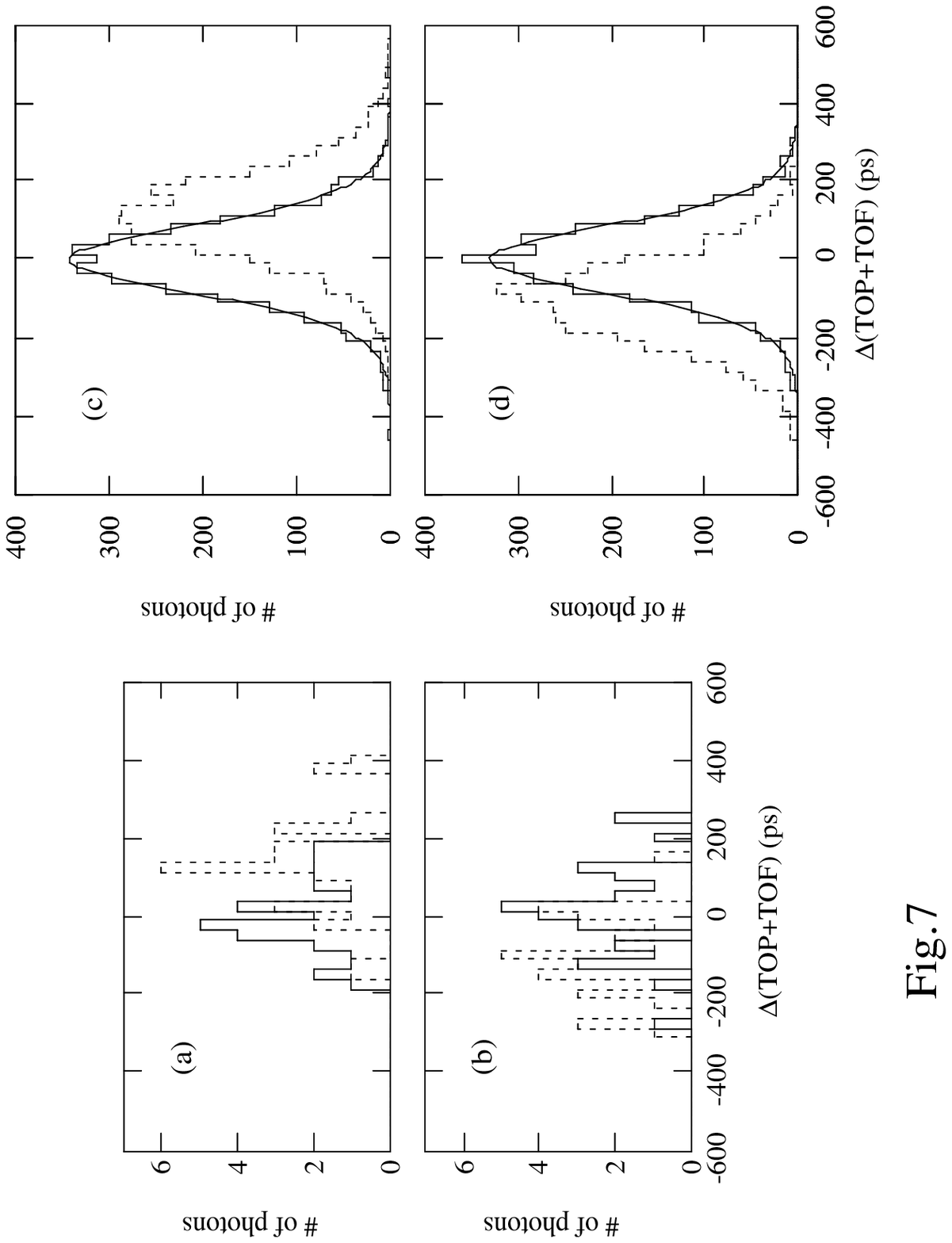}
\end{figure}

\newpage

\begin{figure}
\vspace{24cm}
\hspace{-2cm}
\includegraphics{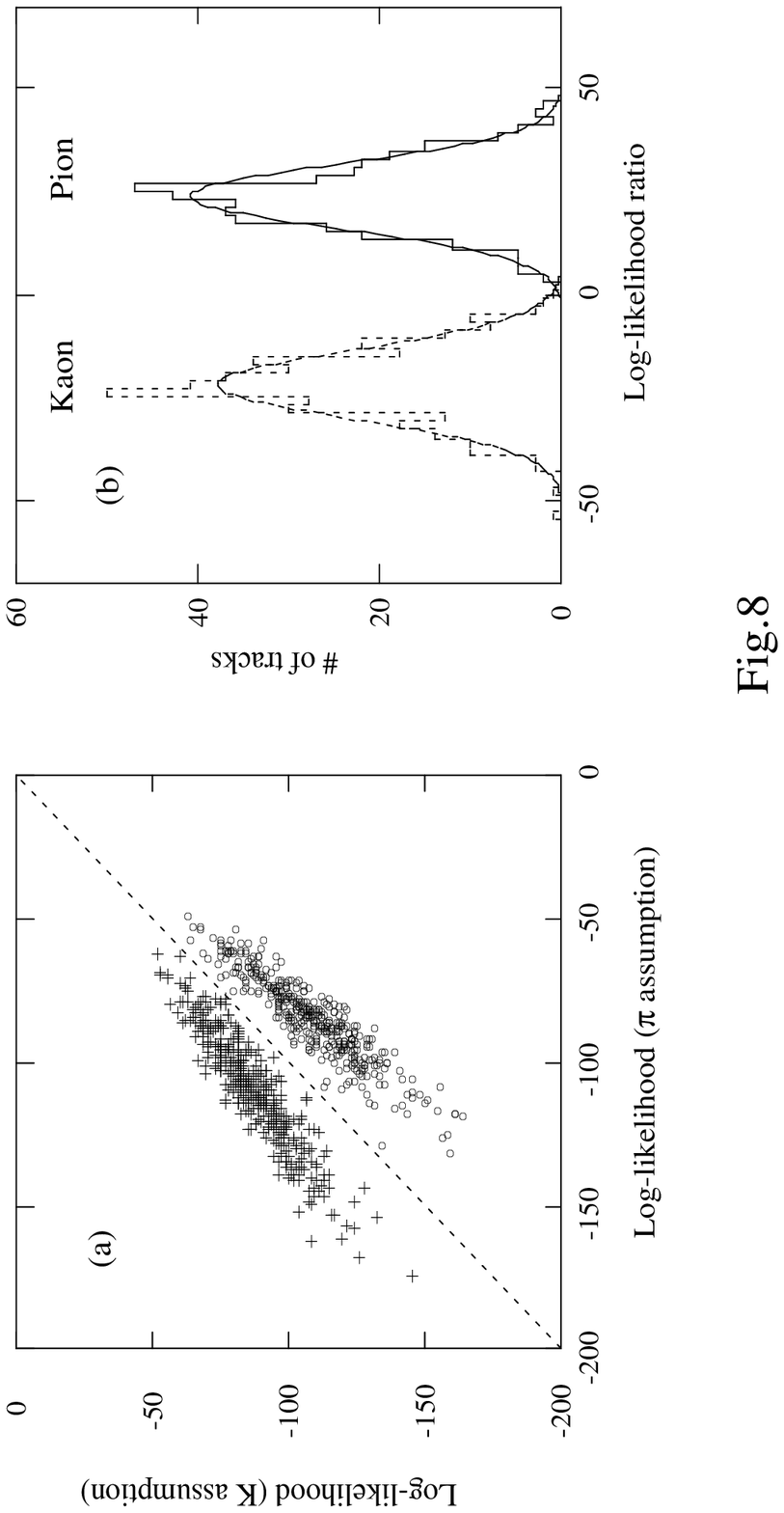}
\end{figure}

\newpage

\begin{figure}
\vspace{24cm}
\hspace{-2cm}
\includegraphics{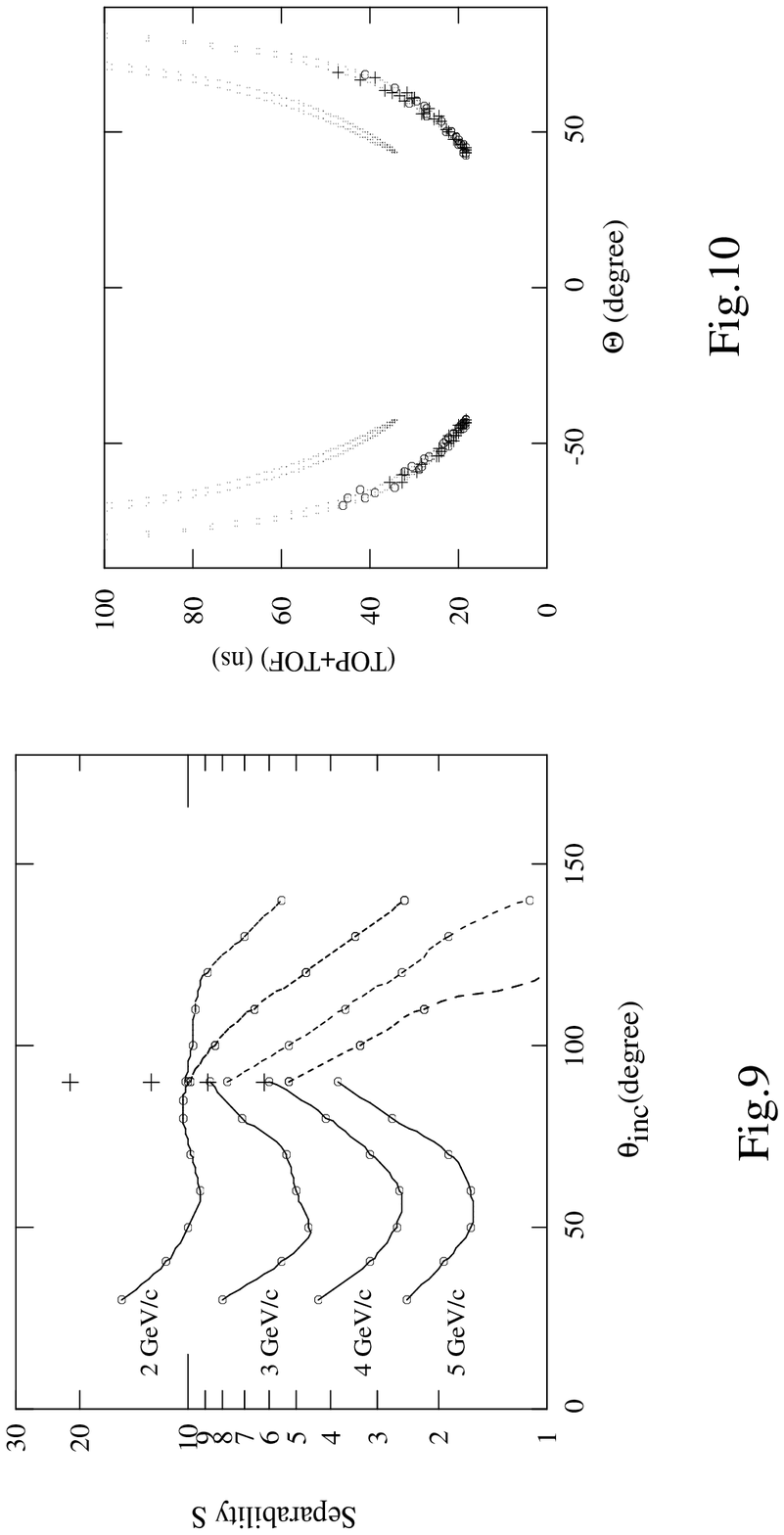}
\end{figure}

\end{description}
\end{document}